\documentstyle[12pt]{article}
\begin{document}
\newcommand{\pn}{\pi^+\pi^-\rightarrow\pi^0\pi^0}
\newcommand{\ph}{\pi^+\pi^-\rightarrow\pi^+\pi^-}
\newcommand{\mn}{m_{\pi^0}}
\newcommand{\mc}{m_{\pi}}
\newcommand{\api}{A\rightarrow\pi^0\pi^0}
\newcommand{\eq}{\begin{equation}}
\newcommand{\en}{\end{equation}}
\newcommand{\ptm}{P^2\to M^2}
\newcommand{\psm}{{P^*}^2\to {M^*}^2}
\newcommand{\phm}{P^2=M^2}
\newcommand{\pqm}{{P^*}^2={M^*}^2}

\begin{titlepage}

\title{
Lifetime of $(\pi^+\pi^-)$ Atom:\\
Analysis of the Role of Strong Interactions.
}

\vspace{0.5cm}

\author{Valery Lyubovitskij\thanks{E-mail: lubovit@thsun1.jinr.dubna.su}
\thanks{On leave from:
Department of Physics, Tomsk State University,
634050 Tomsk, Russia.} \, \, and \,
\, Akaki Rusetsky\thanks{E-mail: rusetsky@thsun1.jinr.dubna.su}
\thanks{On leave from:
IHEP, Tbilisi State University,
University str. 9, 380086 Tbilisi, Georgia.}
\\
\\
Bogoliubov Laboratory of Theoretical Physics, \\
Joint Institute for Nuclear Research,\\
141980 Dubna (Moscow region), Russia}

\maketitle

\vspace{.3cm}

\abstract
{The expression for the $(\pi^+\pi^-)$ atom lifetime is derived
within the Bethe-Salpeter approach. First-order
perturbative corrections due to the contribution
of strong interactions are taken into account.
It is demonstrated that the atom lifetime can be expressed in terms
of the solutions of the Coulombic problem (the wave function of
the $1S$ state at the origin $\Psi_1(0)$, the binding energy of
the $S$-state $E_1$), the difference of the $S$-wave $\pi\pi$
scattering lengths and the energy shift $\Delta E_1$ of the level
due to the strong interactions:
\begin{eqnarray}
\frac{1}{\tau_{1}}\sim (a^0_0-a^2_0)^2|\Psi_1(0)|^2
\biggl(1-\frac{9\Delta E_1}{8E_1}\biggr)\nonumber
\end{eqnarray}
}
\vspace*{.1cm}
\begin{center}
\end{center}
\end{titlepage}

At the present time, the experiment is being prepared by the DIRAC collaboration at
CERN (Experiment PS212) on measurement of the lifetime of
$(\pi^+\pi^-)$ atoms ($A_{2\pi}$) with a 10$\%$ accuracy. The first
estimation of the lifetime of the atom formed by $\pi^+$ and $\pi^-$
in the ground $(1S)$ state $\tau_1 = 2.9^{+\infty}_{-2.1}\times 10^{-15} s$
was obtained in ref.~\cite{Nemenov}.
From a physical point of view, interest in the experiment on
measurement of the $(\pi^+\pi^-)$ atom lifetime stems primarily from
the fact that it allows one to determine the difference of the $S$-wave
$\pi\pi$ scattering lengths $a^0_0-a^2_0$ with the total isospin 0 and 2
in a model-independent way with a 5$\%$ accuracy. The obtained
experimental information about $\pi\pi$ scattering lengths can provide a
decisive test of predictions of the chiral theory \cite{Chiral}.
Recently, high precision experiments on the measurement the characteristics
of both the pionic hydrogen \cite{Sigg} and pionic
deuterium \cite{Chatellard} have also been performed.

For the first time the expression relating $\tau_n$ to the combination
of the $S$-wave hadronic scattering lengths has been obtained in the paper
\cite{Deser}. In this paper
in the framework of the nonrelativistic quantum mechanics the atoms
formed of the proton and $\pi^-$ meson were considered. The main
idea of that paper consisted in the factorization of the strong and
electromagnetic contributions to the width of the $(p\pi^-)$ atom decay
into the pair $n-\pi^0$. Namely, it was assumed that
the energy spectrum of the $(p\pi^-)$ bound state was almost fully determined
by the Coulomb potential as the Borh radius of the atom $r_B \simeq$ 222 fm
was much larger than the strong interaction range. On the other hand,
strong interactions were responsible for the decay of the atom.
In the lowest-order approximation in the fine-structure constant $\alpha$
the atom decay width was written in a form of the product of the square
of the pure Coulombic wave function (w.f.) at the origin and the square
of the difference of the $S$-wave $\pi N$ scattering lengths~\cite{Deser}.
In ref.~\cite{Uretsky} the analogous formula
has been obtained for the case of $\pi^+\pi^-$ atom decay in the
$nS$ state (see \cite{Bilenky} for the corrected expression)
\begin{eqnarray}\label{time}
\frac{1}{\tau_{n}}=
\frac{16\pi}{9}\sqrt{\frac{2\Delta m_\pi}
{m_\pi}}\frac{(a^0_0-a^2_0)^2}{1+2/9m_\pi\Delta m_\pi
(2a^2_0+a^0_0)^2}|\Psi_n(0)|^2
%\frac{1}{64\pi{m_\pi}^2}\sqrt{\frac{2\Delta m_\pi}{m_\pi}}
%|T^{\pn}_{strong}|^2
%|\Psi_n(0)|^2
\end{eqnarray}
\noindent Here the isotopic invariance of pure strong interactions was
assumed, relating the charge exchange amplitude at threshold
$A(\pi^+\pi^-\to \pi^0\pi^0)=(32\pi/3) m_\pi(a^0_0-a^2_0)$
to the scattering lengths in the  $I=0$ and $I=2$ isotopic
channels. The denominator $1+2/9m_\pi\Delta m_\pi (2a^2_0+a^0_0)^2$
in Eq.~(\ref{time}) arises via the unitarization procedure. Further,
$\Psi_n(0)=(m_\pi^3\alpha^3/8\pi n^3)^{1/2}$ is the
nonrelativistic Coulombic w.f. of $A_{2\pi}$ in the $nS$ state
at the origin, $\Delta m_\pi$ is the $\pi^+$-$\pi^0$
meson mass difference and $m_\pi$ is the charged pion mass.

In refs. \cite{Efimov,Pervushin} the strong interaction corrections
to the atom Coulombic w.f. $\Psi_n(0)$ have been estimated, taking into
account the contribution coming only from the discrete spectrum.
It was demonstrated that the strong correction $\Delta\Psi_1(0)$ to
$\Psi_1(0)$ was of an order of $10^{-3}$.
However, if in the calculations the continuous Coulombic spectrum is
taken into account, it is easy to demonstrate that this leads to the
drastic modification of the atom w.f. at the origin. The first-order
perturbative estimate yields the result
$\Delta\Psi_1(0)/\Psi_1(0) \approx (2a_0^0+a_0^2)/(2R) \sim 1/R\geq 35 \%$
assuming that the range of strong potential $R\leq 1$ fm.

Consequently, the inclusion of strong interactions leads to the essential
modification of the Coulombic w.f. of the atom at the origin.
This does not contradict the statement that strong interactions
give a small contribution to the parameters of the $(\pi^+\pi^-)$ bound
state, since the latter implies that the matrix elements of the
strong interaction potential are small compared to the matrix elements
of the Coulombic potential. The w.f., in its turn, is not an
integral characteristic of the system, and it is essentially
modified near the origin where, as is expected from the beginning,
strong interactions should give rise to a sizeable contribution.
On the other hand, it is the Coulombic w.f. that enters into the
expression (\ref{time}) whereas the entire contribution from the
strong interactions is concentrated in the $\pi\pi$ scattering lengths.
Consequently, the inclusion of strong interactions in the atom
w.f. can be regarded as "double counting" and
leads to the erroneous predictions for the $A_{2\pi}$ lifetime.

In the framework of the multichannel potential
theory the strong and electromagnetic corrections to the
observable characteristics of $\pi\pi$ atom have recently been
calculated in ref. \cite{Rasche}. The strong corrections to the
formula (\ref{time}) were calculated in the effective range
approximation (see, also \cite{Trueman}) and
given in a form of the series with an expansion parameter equal
to $A/r_B$ where $A$ denotes the strong $\pi\pi$ scattering amplitude
at threshold.
Since $r_B$ is inversely proportional to the fine structure constant,
these series, in some sense, can be thought to be an expansion in this
constant of the strong amplitude in the presence of Coulombic interaction.
Further, in the chiral theory the $\pi\pi$ scattering amplitude is obtained
in the limit $m_\pi=m_{\pi^0}$. Consequently, for the comparison of
the chiral theory predictions with
the high-precision experimental data it was necessary to evaluate the
effect which stems from the finite $\mc -\mn$ mass difference. In these
calculations as well as during the evaluation of the electromagnetic
corrections the knowledge of the explicit form of strong interaction
potential was required.
Moreover, it turned out that the corrections are rather sensitive to
the particular choice of strong interaction potential \cite{Rasche}.
However, in view of the forthcoming experiment on $A_{2\pi}$
which will provide a consistent test of the chiral theory predictions,
it is necessary to calculate "strong"
corrections directly with the use of the chiral Lagrangian, without any
reference to the concept of phenomenological $\pi\pi$ potential
which is a source of an additional ambiguity in the evaluation
of the atom observables.

On the other hand, in ref. \cite{Silagadze} electromagnetic correction
to the $\pi^+\pi^-$ atom lifetime formula has been calculated.
This correction is caused by the dynamical retardation effect in the
one-photon exchange kernel of Bethe-Salpeter equation for
the atomic wave function. It turned out that this
pure electromagnetic correction ($4\alpha /\pi$) is of the same order of
magnitude as the strong corrections and thus can not be neglected
(in the lowest order these corrections enter additively into the
formula for the atom lifetime).

The aim of the present investigation is to present a self-consistent
field-theoretical framework for the description of the strong decay
of the $\pi^+\pi^-$ atom on the basis of the Bethe-Salpeter (BS) approach.
In this framework, an unambiguous factorization of the strong and
electromagnetic contributions to the expression for the $A_{2\pi}$
lifetime is achieved.
The first-order perturbative corrections due to strong interactions
in the expression for the atom lifetime are calculated without
specifying a concrete form of the strong $\pi\pi$ interaction.

The $(\pi^+\pi^-)$ atom lifetime is calculated according to the well-known
formula \cite{Byckling}
\begin{eqnarray}
\frac{1}{\tau_A}=\frac{\lambda^{1/2}(M^2, m_{\pi^0}^2, m_{\pi^0}^2)}
{32\pi M^3}|T(\api)|^2
\end{eqnarray}
\noindent where $\lambda$ is the well-known kinematic triangle function
and $M=2m_\pi-E_B\approx 2m_\pi$ is the atom mass, $E_B$ being the binding
energy of the $A_{2\pi}$.

After simple transformations we have
\begin{eqnarray}
\frac{1}{\tau_A}=\frac{1}{64\pi m_\pi}
\sqrt{\frac{2\Delta m_\pi}{m_\pi}}
\sqrt{1-\frac{\Delta m_\pi}{2m_\pi}}
|T(\api)|^2
\end{eqnarray}

In the calculations of $T(\api)$, we start from the
standard expression for the transition amplitude for the reactions
involving bound states \cite{Mandelstam,Blankenbecler,Huang}
\begin{eqnarray}\label{init}
T(\api)&=&\lim_{\ptm}i\int\frac{d^4q_1}{(2\pi)^4}
\int \frac{d^4q_2}{(2\pi)^4}
\bar\chi_P(q_1)
\left[ G_0^{-1}(P;q_1,q_2)-V^{\ph}(P;q_1,q_2)\right]\nonumber\\[1mm]
&\times&G^{\pn}(P;q_2,k)
\left[\mn^2-\left(\frac{P}{2}+k\right)^2\right]
\left[\mn^2-\left(\frac{P}{2}-k\right)^2\right]
\end{eqnarray}
\noindent
where $P_{\mu}$ denotes the total 4-momentum of the $\pi^+\pi^-$ atom and
$k_{\mu}$ is the relative 4-momentum of two $\pi^0$ mesons
produced in the decay process.
In the c.m.f. $P_{\mu}=(P_0,\vec 0)$,
$k_0=0,~|\vec k|=\sqrt{P_0^2/4-\mn^2}$.
Here $G_0$ denotes the free Green function of charged $\pi$ mesons,
$G^{\pn}$ is the full Green function for the reaction $\pn$, and $V^{\ph}$
denotes the sum of all irreducible diagrams for the process $\ph$.
The operator $G_0^{-1}-V^{\ph}$
acting on the full Green function $G$,
"excludes" all redundant diagrams which have already been taken into
account in the w.f. $\bar\chi_P$,
thus resolving the "double counting" problem (see, e.g. \cite{Huang}).
The w.f. of the $\pi^+\pi^-$ atom
obeys the BS equation
\begin{eqnarray}\label{BS}
\bar\chi_P(q)G_0^{-1}(P;q)=\int\frac{d^4k}{(2\pi)^4}\bar\chi_P(k)
V^{\ph}(P;k,q),~~~~~~~~P^2=M^2
\end{eqnarray}

The reason why expression (\ref{init}) is not convenient for our
purpose is twofold. First, the w.f. $\bar\chi_P$ contains
the strong interaction contributions. Second, the irreducible kernel
$(G_0^{-1}-V^{\ph})G^{\pn}$ for the transition $\pn$ does not contain
all strong $\pi\pi$ interaction diagrams and, therefore, cannot be
directly related to the experimentally measured charge exchange amplitude.
In order to overcome this difficulty, we transform (\ref{init}) into the
form which is more convenient for further investigations. Namely,
we "transfer" all diagrams corresponding to the strong $\pi^+\pi^-$
interaction from the BS w.f. $\bar\chi_P$ to the irreducible
kernel for the ${\pn}$ transition. To this end, we split the kernel
of Eq. (\ref{BS}) into two parts:
$V^{\ph}(P;k,q)=V_e+V^\prime$ where $V_e$ denotes the instantaneous
Coulombic potential
and $V^\prime$ stands for the rest including, in particular, all strong
interaction diagrams and the piece of one photon exchange diagram
responsible for the dynamical retardation effect.
It should be pointed out that this decomposition is
rather arbitrary; however, for our purpose it is convenient to choose it
in the form given above. From a physical point of view, this corresponds
to a picture in which the instantaneous Coulombic interaction is basically
responsible for the formation of the bound state whereas all other
contributions are small and can be taken into account perturbatively.

Let us now define the new w.f.
\begin{eqnarray}\label{new_BS}
\bar\chi_P(p)=\lim_{{\ptm}\atop{\psm}}
C\int\frac{d^4q_1}{(2\pi)^4}\int \frac{d^4q_2}{(2\pi)^4}
\bar \psi_{P^*}(q_1)\left( G_0^{-1}(P^*;q_1,q_2)-V_e\right)
G^{\ph}(P;q_2,k)
\end{eqnarray}
\noindent Here $M^*=2m_\pi-E_1 + O(\alpha^3)$ is the mass of the bound
state calculated taking into account of only the instantaneous Coulombic
interaction and $C$ is the normalization constant, which will be defined
below. Substituting (\ref{new_BS}) into (\ref{BS}),
it is easy to verify that the new w.f. $\bar\psi_{P^*}$
obeys the BS equation (\ref{BS}) with the displacements
$V^{\ph}\rightarrow V_e$ and $M\rightarrow M^*$. The result
of the action of the operators in (\ref{new_BS}) depends on
the order of limiting procedures. The correct result is obtained
if we assume, e.g.,
$P^2=M^2+\lambda$, ${P^*}^2={M^*}^2+\lambda$, $\lambda\rightarrow 0$.
Substituting the expression (\ref{new_BS}) into (\ref{init}), we get
\begin{eqnarray}\label{Last}
T(\api)&=&\lim_{{\ptm}\atop{\psm}}
iC\int\frac{d^4q_1}{(2\pi)^4}\frac{d^4q_2}{(2\pi)^4}
\bar\psi_{P^*}(q_1)G_e^{-1}(P^*;q_1,q_2)\times\\[1mm]
&\times&G^{\pn}(P;q_2,k)
\left[\mn^2-\left(\frac{P}{2}+k\right)^2\right]
\left[\mn^2-\left(\frac{P}{2}-k\right)^2\right]
\nonumber
\end{eqnarray}

Expression (\ref{Last}) is better suited for our purpose
than (\ref{init}). The irreducible transition kernel entering
into the integrand in (\ref{Last}) contains the total contribution from
strong interactions and $\bar\psi_{P^*}$ is the Coulombic BS w.f.
In the lowest order in $\alpha$ we neglect the difference between
$G_e^{-1}$ and $G_0^{-1}$ and define the strong transition amplitude,
according to the well-known relation
$G^{\pn}_{strong}=G_0^{\pi^+\pi^-}T^{\pn}_{strong}G_0^{\pi^0\pi^0}$.
With the use of the explicit expression for the BS w.f.
$$
\bar\psi_{P^*}(p)=\mc^{-1/2}w({M^*}^2-4w^2)G_0(P^*;p)\tilde\Psi_1(\vec p),
$$
\noindent where $w=\sqrt{m_\pi^2+\vec p^2}$ and the 3-dimensional
Coulombic w.f. in the lowest order in $\alpha$ is written in the
following form \cite{Itzykson}
\begin{eqnarray}\label{function}
\tilde\Psi_1(\vec p)=
\frac{(\alpha\mc)^{3/2}}{(8\pi)^{1/2}}
\frac{4\pi\alpha\mc}{\left(\vec p^2+\mc^2\alpha^2/4\right)^2},
\end{eqnarray}
\noindent we get
$$T(\api)=\lim_{{\ptm}\atop{\psm}}\frac{C}{m_\pi^{1/2}}\hspace*{-.1cm}
\int\frac{d^3\vec q_1}{(2\pi)^3}\tilde\Psi_1(\vec q_1)w({M^*}^2-4w^2)
\hspace*{-.1cm}\int\frac{dq_1^0}{2\pi i}G_0(P;q_1)T^{\pn}_{strong}(P;q_1,k)$$

The w.f. $\tilde\Psi_1(\vec q_1)$ rapidly decreases at the
momenta $\vec q_1^{\,\,2}>\mc^2\alpha^2/4$ so
the main contribution to the integral comes from the area
$|\vec q_1|\approx 0$, where the expression $({M^*}^2-4w^2)$ is small.
In the vicinity of the bound-state pole in the integral over $dq_1^0$,
only the poles of the Green function $G_0(P,q_1)$ can be taken into
account. Integrating over $dq_1^0$ and then over $d^3\vec q_1$,
we get
\begin{eqnarray}
T(\api)&=&C\left. T^{\pn}_{strong}\right|_{thresh.}\mc^{-1/2}
\int\frac{d^3\vec q_1}{(2\pi)^3}\tilde\Psi_1(\vec q_1)
\frac{{M^*}^2-4w^2}{M^2-4w^2}\\[1mm]
&\approx&
C\left. T^{\pn}_{strong}\right|_{thresh.}\mc^{-1/2}\Psi_1(0)(1+\delta)
\nonumber
\end{eqnarray}
\noindent where $\delta = -\Delta E_1/(4E_1)+O(\Delta E_1^2/E_1^2)$,
and $\Psi_1(0)$ is the Schr\"odinger Coulombic w.f. at the origin (\ref{time}) and
$\Delta E_1$ is the energy shift of the 1S Coulombic level ($E_1$)
due to the strong interactions.

The amplitude $T^{\pn}_{strong}$ at the elastic threshold
is expressed through the $S$-wave $\pi\pi$ scattering lengths
(the isotopic invariance of pure strong interactions is assumed when
the scattering amplitudes are expressed in terms of scattering lengths
in $I=0$ and $I=2$ channels)
\begin{eqnarray}
\left. T^{\pn}_{strong}\right|_{thresh.} = \frac{32\pi}{3}\mc
\frac{a^0_0-a^2_0}{1+i(a^0_0+2a^2_0)\sqrt{\frac{2}{9}\mc(\mc-\mn)}}
\nonumber
\end{eqnarray}

The normalization constant $C$ is calculated perturbatively.
To this end, we substitute (\ref{new_BS}) into the normalization
condition for the w.f. $\bar\chi_P(p)$. As a result, we get
\begin{eqnarray}\label{coff}
C=2M^*\left[\int\frac{d^4q_1}{(2\pi)^4}\frac{d^4q_2}{(2\pi)^4}\bar\chi_P(q_1)
\left[\frac{\partial}{\partial P^*_0}G_0^{-1}(P^*;q_1,q_2)\right]
\psi_{P^*}(q_2)\right]^{-1}_{P^*_0=M^*}
\end{eqnarray}
Further, we write the Green function $G^{\ph}$ in Eq. (\ref{new_BS})
in the following form:
\begin{eqnarray}\label{taylor}
G^{\ph}=\left[ G_0^{-1}(P^*)-V_e-V^\prime(P^*)-
(M-M^*)\frac{\partial}{\partial P^*_0}G_0^{-1}(P^*)
+\cdots \right]^{-1}
\end{eqnarray}
\noindent By taking account of (\ref{taylor}), the w.f.
$\bar\chi_P(p)$ in the first perturbative approximation
has the following form
\begin{eqnarray}\label{BS_C}
\bar\chi_P(p)&=&C\lim_{\lambda\to 0}\biggl[\bar\psi_{P^*}(p)+
\int\frac{d^4q_1}{(2\pi)^4}\frac{d^4q_2}{(2\pi)^4}
\bar\psi_{P^*}(q_1)\\[1mm]
&\times&\biggl[V^\prime(P^*;q_1,q_2)-
(M-M^*)\frac{\partial}{\partial P^*_0}G_0^{-1}(P^*;q_1,q_2)\biggr]
G_e(P^*;q_2,p)\biggr]\nonumber
\nonumber
\end{eqnarray}
\noindent where $G_e\equiv G_0^{-1}-V_e$.
Now we multiply Eq. (\ref{BS_C}) by
$\int\frac{d^4k}{(2\pi)^4}\frac{\partial}
{\partial P^*_0}G_0^{-1}(P^*;p,k)\psi_{P^*}(k)$ from the right
and integrate over $d^4p$. Taking account of the normalization condition
for the w.f. $\bar\psi_{P^*}$, the explicit expression
(\ref{coff}) for the constant $C$ and the expression for the energy level
shift in the first perturbative approximation (see, e.g., \cite{Itzykson})
\begin{eqnarray}
2M^*(M-M^*)=\biggl[\int\frac{d^4q_1}{(2\pi)^4}\frac{d^4q_2}{(2\pi)^4}\bar\psi_{P^*}(q_1)
V^\prime(P^*;q_1,q_2)\psi_{P^*}(q_2)\biggr]_{P^*_0=M^*}
\end{eqnarray}
\noindent and after simple transformations we get
\begin{eqnarray}\label{Int}
& &\left[\int\frac{d^4q_1}{(2\pi)^4}\int\frac{d^4q_2}{(2\pi)^4}\bar\chi_P(q_1)
\left[\frac{\partial}{\partial P^*_0}G_0^{-1}(P^*;q_1,q_2)\right]
\psi_{P^*}(q_2)\right]^2_{P^*_0=M^*}\\[1mm]
&=&2M^* \biggl[2M^*+i\int\frac{d^4q_1}{(2\pi)^4}
\frac{d^4q_2}{(2\pi)^4}\bar\psi_{P^*}(q_1)
\biggl[\frac{\partial}{\partial P^*_0}V^\prime
-(M-M^*)\frac{\partial^2}{\partial {P^*_0}^2}G_0^{-1}\biggr]
\psi_{P^*}(q_2)\biggr]_{\pqm}\nonumber
\end{eqnarray}

It is easy to verify that in expression (\ref{Int}) the main
contribution to the integral comes from the term containing
the free Green function
$\frac{\partial^2}{\partial {P^*_0}^2}G_0^{-1}(P^*)$.
Having neglected the dependence of the $\pi^+\pi^-$ strong interaction potential on the
energy in the vicinity of the elastic threshold and calculating the
integral in (\ref{Int}) containing
$\frac{\partial^2}{\partial {P^*_0}^2}G_0^{-1}(P^*)$, we get
\begin{eqnarray}
& &\frac{i}{\mc}\int\frac{d^3\vec q}{(2\pi)^3}\tilde\Psi_1^2(\vec q)
w^2({M^*}^2-4w^2)\int\frac{dq_0}{2\pi i}
G_0(P^*;q)\frac{\partial^2}{\partial {P^*_0}^2}G_0^{-1}(P^*,q)G_0(P^*;q)
\nonumber\\[1mm]
&=&-\frac{10i}{\alpha^2}(1+O(\alpha)) \,\,\,\,\,\Longrightarrow
\,\,\,\,\,
C=1-\frac{5}{16}\frac{\Delta E_1}{E_1}+
O\biggl(\biggl[\frac{\Delta E_1}{E_1}\biggr]^2\biggr)
\nonumber
\end{eqnarray}

Finally, the expression for the $\pi^+\pi^-$ atom lifetime takes
the following form:
$$\frac{1}{\tau_A}=\frac{16\pi}{9}
\sqrt{\frac{2\Delta m_\pi}{m_\pi}}
\sqrt{1-\frac{\Delta m_\pi}{2m_\pi}}
\frac{(a^0_0-a^2_0)^2}{1+\frac{2}{9}\mc(\mc-\mn)(a^0_0+2a^2_0)^2}
\Psi_1^2(0)
\left[1-\underbrace{2\left(\frac{1}{4}+
\frac{5}{16}\right)}_{= 9/8}\frac{\Delta E_1}{E_1}\right]
$$
\noindent where we have separately indicated the corrections coming
from the energy level shift ($-\frac{\Delta E_1}{2E_1}$) and from the
change of the w.f. normalization ($-\frac{5\Delta E_1}{8E_1}$).
Note that the correction due to the change
of the w.f. normalization is a genuine relativistic effect
and arises due to the fact that the free Green function in
the BS equation depends on the bound state mass in a power more than two.
In the quantum mechanics, where $G_0^{-1}(E)=E-H_0$,
$\frac{\partial^2}{\partial E^2}G_0^{-1}(E)=0$ and
the potential does not depend on energy, it is well known that the normalization
of the w.f. does not change in the first order of perturbation theory.

Thus, we have obtained the correction to the formula for the
$\pi^+\pi^-$ atom lifetime \cite{Uretsky} due to strong interactions
in the leading order of the perturbation theory within the field-theoretical
framework based on the Bethe-Salpeter approach. This correction is expressed
in terms of the ratio $\Delta=\Delta E_1/E_1$.
For the estimation of the size of $\Delta$ we use the well-known formula
$\Delta E_1 = (4\pi a_S)/m_\pi \cdot \Psi^2_1(0)$ ~\cite{Deser,Efimov},
relating the energy level shift $\Delta E_1$ to the
$\pi\pi$ scattering singlet length $a_S=2/3a^0_0+1/3a^2_0$.
Consequently, $\Delta=9\Delta E_1/(8E_1) = 9/4 \cdot a_Sm_\pi\alpha
\sim 10^{-3}$ is negligible.
Strong corrections to the $A_{2\pi}$ lifetime formula, obtained in the
present paper are of the same order of magnitude (but have the opposite sign)
as the corrections obtained within the potential picture
\cite{Rasche,Trueman} but not the result from ref. \cite{Geramb}
where an unphysically large value of this correction was obtained.
The small size
of the pure strong first-order corrections indicates that it is important
to evaluate the electromagnetic corrections as well as
to take into account the dynamical retardation effect \cite{Silagadze}
which stems from the noninstantaneous nature of the one-photon exchange
interaction in the 4-dimensional BS approach. These corrections can
perturbatively be taken into account in the irreducible kernel
(\ref{init}), corresponding to the $A_{2\pi}$ decay.

We thank G.V.Efimov, M.A.Ivanov, T.I.Kopaleishvili, E.A.Kuraev,
L.L.Nemenov and A.V. Ta\-ra\-sov for useful discussions, comments
and remarks. This work was supported in part by the Russian Fund of
Fundamental Research (RFFR)  under contract 96-02-17435-a.


\begin{thebibliography}{99}

\bibitem{Nemenov}L.G. Afanasyev et al., Phys.Lett. B 308 (1993) 200;
{\it ibid} B 338 (1994) 478.
\bibitem{Chiral}J.Gasser and H.Leutwyler, Ann.Phys. (N.Y.) 158 (1984) 142;
J.Bijnens et al., NORDITA preprint 95/77 N,P (1995).
\bibitem{Sigg}D. Sigg et al., Phys.Rev.Lett. 75 (1995) 3245.
\bibitem{Chatellard}D. Chatellard et al., Phys.Rev.Lett. 74 (1995) 4157.
\bibitem{Deser}S. Deser et al., Phys.Rev. 96 (1954) 774.
\bibitem{Uretsky}J.L. Uretsky and T.R. Palfrey, Phys.Rev. 121 (1961) 1798.
\bibitem{Bilenky}S.M. Bilenky et al., Yad.Fiz. 10 (1969) 812.
\bibitem{Efimov}G.V. Efimov, M.A. Ivanov and V.E. Lyubovitskij,
Sov.Jour.Nucl.Phys. 44 (1986) 296.
\bibitem{Pervushin}A.A. Bel'kov, V.N. Pervushin and F.G. Tkebuchava,
Sov.Jour.Nucl.Phys. 44 (1986) 300.
\bibitem{Rasche}U. Moor, G. Rasche, W.S. Woolcock,
Nucl.Phys. A 587 (1995) 747.
\bibitem{Trueman}T.L.Trueman, Nucl.Phys. 26 (1961) 57.
\bibitem{Silagadze}Z. Silagadze, JETP Lett. 60 (1994) 689.
\bibitem{Byckling}E. Byckling and K. Kajantie,
Particle Kinematics (John Wiley and Sons, 1973).
\bibitem{Mandelstam}S. Mandelstam, Proc.Roy.Soc. 223 (1955) 248;
\bibitem{Blankenbecler}R. Blankenbecler, Nucl.Phys. 14 (1959/1960) 97.
\bibitem{Huang}K. Huang and H.A. Weldon, Phys.Rev. D 11 (1975) 257.
\bibitem{Itzykson}C. Itzykson and J.B. Zuber, Quantum Field Theory
(McGraw-Hill, 1980).
\bibitem{Geramb}M. Sander, C. Kuhrts and H.V. von Geramb,
Phys.Rev. C 53 (1996) R2610.
\end{thebibliography}
\end{document}